\documentclass[10pt]{article}
\usepackage[french,english]{babel}
\usepackage{amsmath}
\usepackage{amsfonts}
\usepackage{amssymb}
\usepackage{graphicx}  
\usepackage{times}
\def\psl{/\ \hskip -9pt p}

\begin{document}

%***********************************************************************
%
 April 30th 2016   \hfill %  IJMPA-revised.tex
%\hskip 2cm re-submitted to IJMPA on March 25th 2016

to appear in IJMP\;A.\quad DOI: 10.1142/S0217751X16500718\hfill

%DOI: 10.1142/S0217751X16500718\hfill accepted for publication on April 7th 2016
%\today\quad   \hfill IJMPA-revised.tex
\vskip 5cm
{\baselineskip 12pt
\begin{center}
{\bf THE 1-LOOP SELF-ENERGY OF AN ELECTRON

IN A STRONG EXTERNAL MAGNETIC FIELD  REVISITED}
\end{center}
}
\baselineskip 16pt
\arraycolsep 3pt  
\vskip .2cm
\centerline{
B.~Machet
     \footnote{Sorbonne Universit\'es, UPMC Univ Paris 06, UMR 7589,
LPTHE, F-75005, Paris, France}
     \footnote{CNRS, UMR 7589, LPTHE, F-75005, Paris, France.}
     \footnote{Postal address:
LPTHE tour 13-14, 4\raise 3pt \hbox{\tiny \`eme} \'etage,
          UPMC Univ Paris 06, BP 126, 4 place Jussieu,
          F-75252 Paris Cedex 05 (France)}
    \footnote{machet@lpthe.jussieu.fr}
     }
\vskip 1cm

{\bf Abstract:} I calculate the 1-loop self-energy of the lowest Landau
level of an electron of mass $m$ in a strong, constant and uniform
external magnetic field $B$,
beyond its always used truncation at $(\ln L)^2, L=\frac{|e|B}{m^2}$.
This is achieved by evaluating the integral deduced in 1953 by Demeur and
incompletely calculated in 1969 by Jancovici, which I recover from
Schwinger's techniques of calculation.  It yields
$\delta m \simeq \displaystyle\frac{\alpha m}{4\pi}
\Bigg[\left(\ln L - \gamma_E - \frac32\right)^2 - \frac94  + \frac{\pi}{\beta-1}
+ \frac{\pi^2}{6} + \frac{\pi\;\Gamma[1-\beta]}{L^{\beta-1}}
+ \frac{1}{L}\left(\frac{\pi}{2-\beta} - 5\right)
+{\cal O}(\frac{1}{L^{\geq 2}})\Bigg]$ with $\beta \simeq 1.175$ for
$75 \leq L \leq 10\,000$.
The  $(\ln L)^2$ truncation exceeds the precise estimate
by $45\%$ at $L=100$   and by more at lower values of $L$,
due to  neglecting, among others, the single logarithmic contribution.
This is doubly unjustified because it is large and because it is needed to
fulfill appropriate renormalization conditions.
Technically challenging improvements look therefore necessary, for example
when resumming higher loops and incorporating the effects of large $B$
on the photonic vacuum polarization, like investigated in recent years.

\bigskip


\newpage
%
%SSSSSSSSSSSSSSSSSSSSSSSSSSSSSSSSSSSSSSSSSSSSSSSSSSSSSSSSSSSSSSSS
\section{Generalities}
%SSSSSSSSSSSSSSSSSSSSSSSSSSSSSSSSSSSSSSSSSSSSSSSSSSSSSSSSSSSSSSSS

We shall be concerned in this short 
\footnote{This is why I do not pay a fair enough tribute  to the
many authors that contributed to this subject, and I apologize for this.
I will instead insist on very small details, generally not
mentioned, that can help the reader.} 
note, with the self-energy of an electron at 1-loop in the presence of a
strong, constant and uniform external magnetic field $B$.
The electron propagator is described by the sum of the 2 diagrams of Fig.~1
%
%ffffffffffffffffffffffffff FIGURE 1 self.pdf fffffffffffffffffffffffffffffff
\begin{center}
\includegraphics[width=12cm, height=1.5cm]{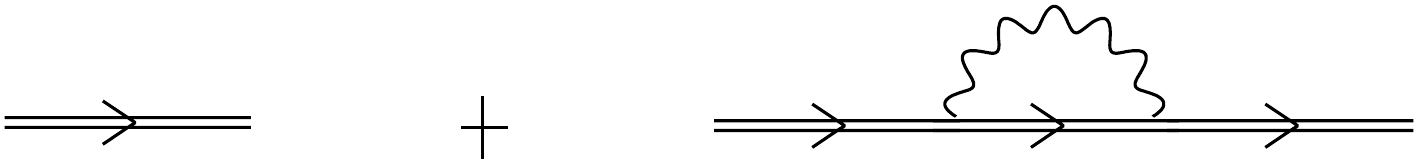}

{\em Fig.~1: 1-loop radiative correction to the mass of an electron.}
\end{center}
%ffffffffffffffffffffffffff FIGURE 1 self.pdf fffffffffffffffffffffffffffffff
%
in which the double horizontal lines, external as well as internal,
stand for  an electron of mass $m$ in an external $B$.
The electron mass is defined as the pole of its propagator, which is the
only gauge invariant definition. Renormalization conditions are set
accordingly.
The self-energy that we shall calculate is the second diagram.
For the sake of simplicity, we shall restrict external electrons to lie
in the lowest Landau level. This  does not apply to the
internal electron propagator, which includes a summation on all Landau
levels.

\subsection{Motivation}
%sssssssssssssssssssssss

The uses of the self-energy of an electron in a strong
external $B$ generally rely  on its  ``leading'' double logarithmic term 
proportional to $\left(\ln\displaystyle\frac{|e|B}{m^2}\right)^2$ and  its eventual
transmutation into a single logarithmic behavior on accounting for the
accompanying modifications of the photonic vacuum polarization.
The double logarithmic term was first extracted in 1969 by Jancovici
\cite{Jancovici}  from a general formula deduced by Demeur in 1953
\cite{Demeur} \footnote{ As far as I could see, Demeur's calculations,
performed with techniques which are unfamiliar today,
have not been reproduced.  They have been critically examined and completed
by Newton \cite{Newton} at small
values of $\frac{|e|B}{m^2}$, but this path seems to have then been
abandoned.}. At the very end of \cite{Jancovici}, Jancovici mentions the
presence of potentially large, single logarithmic and constant corrections,
but the constant $A$ could not be determined at that time.
The asymptotic double logarithmic behavior  at $B\to \infty$
was also obtained by Loskutov and Skobelev in 1977 \cite{LS1977}
\footnote{see also \cite{LS1979}.}
 in the 2-dimensional limit of QED which is  a  suitable approximation  at this
limit. However, only the kinematical
domains of integrations leading to the double logs were accounted for and
the eventual presence of large but non
double-logarithmic corrections was not investigated.
The next  step is a resummation of the same double-logarithmic terms 
in rainbow-type diagrams  argued to be dominant. In \cite{LS1981}, Loskutov
and Skobelev have shown in 1981 that the result exponentiates.
A slightly different result, non-exponential, was obtained later in 1999
by Gusynin and Smilga in \cite{GusyninSmilga}, who were still only concerned by
resuming double logarithmic terms.
An important modification to be brought to these results had already been
shown earlier in 1983 in \cite{LS1983}  again
by Loskutov and Skobelev, then studied more extensively 
in 2002 in \cite{KMO-MPLA-2002} by Kuznetsov, Mikheev and Osipov:
accounting for the effective photon mass induced by asymptotically strong
magnetic fields shrinks the double logarithm down
to a single logarithmic behavior. While in \cite{LS1983} an exponentiation
still occurs, a different result is obtained in \cite{KMO-MPLA-2002} in which
higher Landau levels for the virtual electron are also included.

In all these calculations, only double
logarithmic terms were considered at the start, which would then be
eventually resummed and corrected by an effective photon mass.
This makes that the corresponding results never incorporate the starting large
single logarithm and constant  which, as we shall show,
strongly damp the $\ln^2$ truncation  of the 1-loop electron
self-energy.
This neglect is all the more unfortunate as  the large single logarithm
is tightly connected to renormalization conditions and to the corresponding
counterterms.
Getting meaningful results requires indeed that
the appropriate renormalization conditions should be fulfilled at each
order of the resummation process, 
and that the same care be due when including
the corrections to the photon vacuum polarization by the external
$B$.
Calculations as they have been done up to now, that gave birth to
many developments, for example in condensed matter physics
\footnote{see for example the review \cite{MiranskyShovkovy}.}
,
do not seem to worry about  these criteria, which can jeopardize
their conclusions and predictions.

\subsection{The procedure}
%sssssssssssssssssssssssss

I will re-calculate the Demeur-Jancovici
integral and explicitly display the large corrections that strongly damp
the double logarithmic behavior of the electron self-energy in a strong
external $B$. Special attention will be paid to the counterterm that
ensures suitable renormalization conditions for the electron self-energy.

To cast this on solid grounds, I will first show, in section
\ref{section:equiv},  how this integral
can be recovered by using the formalism developed by Schwinger in the late 1940's
\cite{Schwinger1949-2}.
 The corresponding calculations are explained in details in the book by
Dittrich and Reuter \cite{DittrichReuter} in 1985 (which includes
a long list of references). One finds in there, in particular, the expression
for the renormalized 1-loop mass  operator $\Sigma(\pi)$, where $\pi_\mu =
p_\mu -eA_\mu$,
 for an electron in an external $B$, as deduced in 1974 by Tsai
\cite{Tsai1974}
\footnote{ At the end of his paper, Tsai just states
that his calculation, which uses the techniques and results of Schwinger,
yields, when projected on the ground state of the
electron, ``...the known result of Demeur'' (this correspondence is the
subject of subsection \ref{subsec:changevar}).}.
It  will be the  starting point of the  original calculations.

I will then make then use of Demeur's technique
\cite{Demeur} to sandwich the mass operator $\Sigma(\pi)$ 
 between two ``privileged'' electron states $|\;\psi>$ 
 (to reproduce the terminology of Demeur  and
previous authors, in particular Luttinger \cite{Luttinger}),
on mass-shell.
This restricts, but greatly simplifies the calculations.
This matrix element corresponds to $\delta
m$ of the electron at 1-loop in the presence of $B$.
The privileged state, that
always exists in the presence of $B$, is the one with energy $m$.
In our present terminology, it corresponds to the
Lowest Landau Level (LLL) and, on mass shell, it satisfies the
Dirac equation $(\pi\!\!\!/ +m)|\;\psi>=0$
\footnote{I use  Schwinger's metric $(-,+,+,+)$.}.

Then, I will  show how changes of variables cast $\delta m$  in the
form deduced by Demeur \cite{Demeur} and used by Jancovici \cite{Jancovici}.
It is a convergent double integral that only depends
on $\displaystyle\frac{|e|B}{m^2}$.
Its rigorous exact analytical evaluation lies  beyond my
ability. However, a trick due to  M.I.~Vysotsky 
in his study of the screening of the Coulomb potential in an external magnetic
field \cite{Vysotsky2010} comes to the rescue:
the part of the integrand that resists analytical
integration can be nearly perfectly fitted inside the range of integration
by a simpler function that can be analytically integrated.

%SSSSSSSSSSSSSSSSSSSSSSSSSSSSSSSSSSSSSSSSSSSSSSSSSSSSSSSSSSSSSSSSSSSSSSSSS
\section{The 1-loop self-energy $\boldsymbol \Sigma$ for the lowest Landau
level of an electron in external $\boldsymbol{B}$;
equivalence between the calculations by Schwinger and Demeur}
\label{section:equiv}
%SSSSSSSSSSSSSSSSSSSSSSSSSSSSSSSSSSSSSSSSSSSSSSSSSSSSSSSSSSSSSSSSSSSSSSSSS

\subsection{The general formula for the electron self-energy operator
at 1-loop}\label{subsec:genform}
%sssssssssssssssssssssssssssssssssssssssssssssssssssssssssssssssssssss

For this work to be self-contained, I  recall the main steps in the
determination of the operatorial expression of the
self-energy of an electron in an external $B$ deduced by
Tsai \cite{Tsai1974}. I closely follow the book by Dittrich
and Reuter \cite{DittrichReuter}, more precisely  the
paragraphs 2 and 3, that I summarize here. It means
that the present subsection does not include anything original and owes everything to
\cite{DittrichReuter}.

The self-energy $\Sigma(x',x'')$ includes 2 internal propagators:\newline
*\ the free photon propagator, that we shall take in the Feynman gauge, arguing
of the gauge independence of Schwinger's techniques of calculation
\cite{Schwinger1949-2}\cite{Schwinger1951}
\begin{equation}
i\Delta_{\mu\nu}(x'-x'') = -ig_{\mu\nu} D(x'-x''),\quad
D(x'-x'')= \int\frac{d^4k}{(2\pi)^4}\;
e^{ik(x'-x'')}\;\frac{1}{k^2-i\epsilon}.
\end{equation}
*\ the electron propagator in the presence of an external magnetic field
as determined by Schwinger \cite{Schwinger1951} and then re-expressed and
used by Tsai \cite{Tsai1974} to make the link with the calculations by Demeur
\cite{Demeur}
\begin{equation}
G(x',x'',B)\equiv i<0\;|\;T\;\psi(x')\bar\psi(x'')\;|\;0>
=\Phi(x',x'')\int\frac{d^4p}{(2\pi)^4}\;e^{ip(x'-x'')}\;G(p,B),
\end{equation}
in which the phase $\Phi(x',x'')$, which ensures gauge invariance, is given by
\begin{equation}
\Phi(x',x'')= \exp\Bigg[ie\int_{x''}^{x'} dx_\mu\Big(A^\mu(x)+\frac12
F^{\mu\nu}(x'_\nu -x''_\nu)\Big)
\Bigg],
\label{eq:phase}
\end{equation}
and (this is eq.(2.47b) of \cite{DittrichReuter})
%\footnote{$e$ stands here for the charge of the  electron $e=-|e|<0$.}
\begin{equation}
G(p, B) =i\int_0^\infty ds_1\; e^{-is_1\big(m^2-i\epsilon+ p_\parallel^2
+\frac{\tan z}{z}\;p_\perp^2\big)}\;\frac{e^{i z\sigma_3}}{\cos z}
\left(m-p\!\!\!/_\parallel-\frac{e^{-i z\sigma_3}}
{\cos z}\;p\!\!\!/_\perp \right),\quad with\ z = eBs_1.
\label{eq:eprop}
\end{equation}
$e$ stands everywhere in this work for the charge of the  electron $e=-|e|<0$.
The metric that is used is $(-1,+1,+1,+1)$ and the notations are the following
\begin{equation}
\begin{split}
& \sigma^3 =\sigma^{12}=\frac{i}{2}[\gamma^1,\gamma^2]=
diag(1,-1,1,-1),\cr
& p_\parallel=(p_0,0,0,p_3),\quad p_\parallel^2 = -p_0^2 + p_3^2,\quad
p_\perp=(0,p_1,p_2,0),\quad p_\perp^2 = p_1^2 + p_2^2,\cr
& p\!\!\!/_\parallel=-\gamma_0 p_0 + \gamma_3 p_3,\quad
p\!\!\!/_\perp=\gamma_1 p_1 + \gamma_2 p_2.
\end{split}
\label{eq:notations}
\end{equation}
The propagator (\ref{eq:eprop}) includes all Landau levels of the electron.

The constant external $B$ is chosen in the $z$-direction
such that $F_{12}=-F_{21}=B$ (therefore the notations ``$\parallel$'' and
``$\perp$'' have a natural meaning).

The phase (\ref{eq:phase}) is independent of the choice of the path of
integration because the curl of the integrand vanishes. Choosing a straight
line of integration $x(t)=x''+t(x'-x''), t\in[0,1]$
 leads to the familiar expression
\begin{equation}
\Phi(x',x'')=e^{ie\int_{x''}^{x'}dx_\mu\,A^\mu(x)}.
\end{equation}
\subsubsection{The unrenormalized self-energy}
%=============================================
%
$\bullet$\ In terms of the quantities above, the 1-loop self-energy writes
(``c.t.'' stands for ``counterterms'')
\begin{equation}
\Sigma(x',x'',B)= ie^2 \gamma^\mu\; G(x',x'',B)\,D(x'-x'')\;\gamma_\mu + c.t.
\end{equation}
that is
\begin{equation}
\begin{split}
\Sigma(x',x'',B) &=\Phi(x',x'')
\int\frac{d^4p}{(2\pi)^4}\;e^{ip(x'-x'')}\;\Sigma(p,B),\cr
\Sigma(p,B) &=ie^2\gamma^\mu
\int\frac{d^4k}{(2\pi)^4}\;\frac{1}{k^2-i\epsilon}\;G(p-k,B)\;\gamma_\mu
+c.t.
\end{split}
\end{equation}
One introduces a second Schwinger parameter $s_2$ for the photon propagator
\begin{equation}
\frac{1}{k^2-i\epsilon}= i\int_0^\infty ds_2\;e^{-is_2(k^2-i\epsilon)},
\end{equation}
and get eq.~(3.11) of \cite{DittrichReuter}
\begin{equation}
\begin{split}
\Sigma(p,B) &= -ie^2 \int_0^\infty ds_1 \int_0^\infty ds_2 \int\frac{d^4
k}{(2\pi)^4}\;e^{-is_2(k^2-i\epsilon)}\;
e^{-is_1\big(m^2+(p-k)_\parallel^2 +\frac{\tan z}{z}\;(p-k)_\perp^2
\big)}\cr
& \hskip 3cm \gamma^\mu\;\frac{e^{i z\sigma^3}}{\cos z}
\Big( m-(p\!\!\!/_\parallel -k\!\!\!/_\parallel)-\frac{e^{-iz\sigma^3}}
{\cos z}\; (p\!\!\!/_\perp -k\!\!\!/_\perp)
\Big)\;\gamma_\mu + c.t.
\end{split}
\end{equation}
$\bullet$\ The next step is  to change variables: one goes
from $s_1$ and $s_2$ to $s$ and $u$ such that
\begin{equation}
s_1=su,\quad s_2=s(1-u) \Rightarrow
\int_0^\infty ds_1 \int_0^\infty ds_2 = \int_0^\infty ds\; s \int_0^1 du,\quad
Y=eBsu.
\end{equation}
$\Sigma(p,B)$ can then be cast in the form
\begin{equation}
\begin{split}
\Sigma(p,B) &= -ie^2 \int_0^\infty ds\; s\int_0^1 
\frac{du}{\cos Y}\; \Big\{ \int\frac{d^4k}{(2\pi)^4}\; e^{-is\chi}\Big\}
\gamma^\mu\;e^{iY\sigma^3}
\Bigg[m-(1-u)\,p\!\!\!/_\parallel +\frac{e^{-iY\sigma^3}}{\cos Y}\;
\frac{1-u}{1-u+u\,\displaystyle\frac{\tan Y}{Y}}\;p\!\!\!/_\perp \Bigg]\gamma_\mu +c.t.\cr
\chi &= um^2 + \varphi+(k_\parallel-up_\parallel)^2 +\Big(1-u+u\,\displaystyle\frac{\tan Y}{Y}\Big)
\Bigg(k_\perp-\frac{u\,\displaystyle\frac{\tan Y}{Y}}{1-u+u\,\displaystyle\frac{\tan
Y}{Y}}\;p_\perp\Bigg)^2,\cr
\varphi &= u(1-u)\,p_\parallel^2 +\frac{u}{Y}\;\frac{(1-u)\sin Y}{(1-u)\cos Y
+u\,\displaystyle\frac{\sin Y}{Y}}\;p_\perp^2.
\label{eq:sig4}
\end{split}
\end{equation}
The $k$ integration, which only occurs inside the curly bracket in
(\ref{eq:sig4}) can now be performed by shifting the integration
variables inside $\chi$ and by using the standard integral
$\int_{-\infty}^{+\infty} dx\;e^{\pm i A x^2}=e^{\pm
i\frac{\pi}{4}}\;\left(\frac{\pi}{A}\right)^{1/2}, A>0$, which yields
eq.~(3.27) of \cite{DittrichReuter} ($\alpha=\displaystyle\frac{e^2}{4\pi}$)
\begin{equation}
\begin{split}
\Sigma(p,B) &=\frac{\alpha m}{2\pi}\int_0^\infty\frac{ds}{s}\int_0^1 du\;
\frac{e^{-is(um^2+\varphi)}}{(1-u)\cos Y+u\,\displaystyle\frac{\sin Y}{Y}}\;e^{iY\sigma^3}
\Bigg[1+e^{-2iY\sigma^3} \cr
& \hskip 4cm+(1-u)\,e^{-2iY\sigma^3}\;\frac{p\!\!\!/_\parallel}{m} +(1-u)\;\frac{e^{-iY\sigma^3}}{(1-u)\cos
Y + u\,\displaystyle\frac{\sin Y}{Y}}\;\frac{p\!\!\!/_\perp}{m}
\Bigg] +c.t.
\end{split}
\end{equation}
At this stage, the integrations on $s$ and $u$ cannot be done explicitly.

$\bullet$\ The last and crucial step to get the self-mass $\delta m$ of a
given state $|\;\psi>$ on mass-shell ($(\pi\!\!\!/+m)|\;\psi>=0$)
is  to go to the so-called ``space
representation'' and $\Sigma(\pi)$ defined by
\begin{equation}
\Sigma(x',x'',B) = \Phi(x',x'')
\int\frac{d^4p}{(2\pi)^4}\;e^{-ip(x'-x'')}\;\Sigma(p,B)
=<x'\;|\; \Sigma(\pi)\;|\;x''>.
\label{eq:spacerep}
\end{equation}
Note that the phase $\Phi(x',x'')$ gets now ``included'' in $\Sigma(\pi)$.

For this, one has to go through the manipulations of pages 47-50 of
\cite{DittrichReuter}. We shall only write here the intermediate
formul{\ae} (eventually correcting for some misprints).
One is led to introduce
\begin{equation}
\Delta= (1-u)^2 +2u(1-u)\cos Y \frac{\sin Y}{Y} + u^2 \left(\frac{\sin
Y}{Y}\right)^2,
\label{eq:Delta}
\end{equation}
the angle $\beta$ such that
\footnote{In eq.~(3.31) of \cite{DittrichReuter}, the first 2 expressions
for $\cos\beta$ should be replaced by their inverse.}
\begin{equation}
\cos\beta =\frac{(1-u)\cos Y +u\,\displaystyle\frac{\sin Y}{Y}}{\Delta^{1/2}},\quad
\sin \beta=\frac{(1-u)\sin Y}{\Delta^{1/2}},
\label{eq:beta}
\end{equation}
and
\footnote{There is a sign misprint in the definition (3.38b) of
$\Phi$ in \cite{DittrichReuter}, which has been corrected here. The
correct sign is the one in eq.~(3.35) of
\cite{DittrichReuter}.\label{foot:sign}}
\footnote{This $\Phi$ should not be confused with the phase $\Phi(x',x'')$
of (\ref{eq:phase}).}
\begin{equation}
\Phi= u(1-u)\big(m^2 - \pi\!\!\!/^2\big)+\frac{u}{Y} \big(
\beta -(1-u)Y\big) \pi_\perp^2 -u^2\, \frac{e}{2}\,\sigma_{\mu\nu}\,F^{\mu\nu}.
\label{eq:Phi}
\end{equation}
One gets then
\begin{equation}
\begin{split}
\Sigma(\pi) &= \frac{\alpha m}{2\pi} \int_0^\infty\frac{ds}{s}\int_0^1 du\;
e^{-isu^2m^2}\Bigg[
\frac{e^{-is\Phi}}{\Delta^{1/2}} \Big[
1+ e^{-2iY\sigma^3}+(1-u)\,e^{-2iY\sigma^3}\;\frac{\pi\!\!\!/}{m}\cr
& \hskip 2cm +(1-u)\Big(\frac{1-u}{\Delta}+\frac{u}{\Delta}\;\frac{\sin Y}{Y}
\;e^{-iY\sigma^3}-e^{-2iY\sigma^3}\Big)\frac{\pi\!\!\!/_\perp}{m}
\Big]
+c.t.\Bigg].
\end{split}
\end{equation}

\subsubsection{The renormalization conditions and the counterterms}
\label{subsub:rencond}
%==================================================================

Since we are working in a gauge invariant formalism,  consistency requires
that the
mass of the electron be also defined in a gauge invariant way, that is as
the pole of its propagator.

*\ At $B=0$, the electron propagator writes
\begin{equation}
G(p) = \frac{1}{p\!\!\!/ + m_0 + \Sigma(p)},
\end{equation}
in which $\Sigma(p)$ is the bare quantity,
 and the renormalized electron mass is accordingly defined by
\begin{equation}
m=m_0+\delta m,\quad \delta m = \Sigma(p)_{p\!\!\!/+m=0}.
\label{eq:defmass}
\end{equation}

*\ At $B\not=0$, the electron propagator is
\begin{equation}
G= \frac{1}{\pi\!\!\!/ +m_0 + \Sigma(\pi)}.
\end{equation}
We define, in analogy with eq.~(\ref{eq:defmass}),
the mass of the electron as the pole of its propagator by
\begin{equation}
m=m_0+\Sigma(\pi)_{\pi\!\!\!/+m=0} \Leftrightarrow \delta m
=\Sigma(\pi)_{\pi\!\!\!/+m=0}.
\label{eq:mdef}
\end{equation}
$\delta m$ depends on the external field. Note that, on mass-shell,
$\pi\!\!\!/^2 \equiv  -\pi^2
+\frac{e}{2}\,\sigma_{\mu\nu}F^{\mu\nu}= m^2$.

The counterterms
 are determined by the two equations (3.39) and (3.40) of
\cite{DittrichReuter} (we restore the superscript $``ren''$ to  make clear
that
one deals now with the renormalized quantities)
\footnote{These renormalization
conditions are carefully explained in p.~38-41 of \cite{DittrichReuter}.
Their importance is also emphasized by Ritus in \cite{Ritus1975}.}
\begin{equation}
 \lim_{\pi\!\!\!/ +m=0}\lim_{B\to 0} \Sigma^{ren}(\pi)=0,\quad
 \lim_{\pi\!\!\!/ +m=0}\lim_{B\to 0}
\frac{\partial\Sigma^{ren}(\pi)}{\partial\pi\!\!\!/}=0.
\label{eq:renorm}
\end{equation}
They ensure that, after turning off $B$, ($\pi\!\!\!/ \to \psl$),
 the renormalization
conditions $\Sigma^{ren}(p)_{p\!\!\!/+m=0}=0$ and
$\displaystyle\frac{\partial\Sigma^{ren}(p)}{\partial
p\!\!\!/}\Big|_{p\!\!\!/+m=0}=0$ are fulfilled (compare with
(\ref{eq:defmass})).

Since the renormalization conditions are expressed at $B=0$, one needs the
following limits at $Y\equiv eBsu \to 0$
\begin{equation}
\begin{split}
& \beta \stackrel{Y\to 0}{\to} (1-u)Y + {\cal O}(Y^2),\cr
& \Phi \stackrel{Y\to 0}{\to} u(1-u)
(m^2-\pi\!\!\!/^2)=u(1-u)(m^2-p\!\!\!/^2),\cr
& \Delta \stackrel{Y\to 0}{\to} 1.
\end{split}
\end{equation}
At $B=0$, $\pi = p$, and one gets
\begin{equation}
\Sigma(p) =\frac{\alpha m}{2\pi}\int_0^\infty \frac{ds}{s}\int_0^1 du\;
e^{-isu^2m^2}\Big[e^{-isu(1-u)}\;(m^2-p\!\!\!/^2)\big(2+(1-u)
\frac{p\!\!\!/}{m}\big)+c.t.\Big],
\end{equation}
and, at $p\!\!\!/+m=0$, point at which the renormalization conditions are
expressed
\begin{equation}
\Sigma(p)\big|_{p\!\!\!/+m=0}=\frac{\alpha
m}{2\pi}\int_0^\infty\frac{ds}{s}\int_0^1 du\;
e^{-isu^2m^2}\;\big[(1+u)+c.t.\big].
\end{equation}
To fulfill the first renormalization condition, we must therefore introduce
a first counterterm
\begin{equation}
c.t._1= -(1+u).
\end{equation}
To implement the second renormalization condition, one calculates
\begin{equation}
\hskip -1cm
\frac{\partial \Sigma(p)}{\partial p\!\!\!/} = \frac{\alpha m}{2\pi}
\int_0^\infty \frac{ds}{s} \int_0^1 du\;e^{-isu^2m^2}\Bigg[
(-is)u(1-u)(-2p\!\!\!/)e^{-isu(1-u)(m^2-p\!\!\!/^2)}\;\bigg(2+(1-u)\frac{p\!\!\!/}{m}\bigg)
+\frac{1-u}{m}\;e^{-isu(1-u)(m^2-p\!\!\!/^2)}\Bigg],
\end{equation}
such that, at $p\!\!\!/+m=0$ one gets
\begin{equation}
\frac{\partial \Sigma(p)}{\partial
p\!\!\!/}\Big|_{p\!\!\!/+m=0}=\frac{\alpha
m}{2\pi}\int_0^\infty\frac{ds}{s}\int_0^1
du\;e^{-isu^2m^2}\Big[-2ismu(1-u^2)+\frac{1-u}{m}\Big].
\end{equation}
This leads to the second counterterm
\begin{equation}
c.t._2= -(\pi\!\!\!/+m)\Big(\frac{1-u}{m}-2imu(1-u^2)s\Big),
\end{equation}
in which the factor $(\pi\!\!\!/+m)$ ensures that the first
renormalization condition keeps satisfied.

\subsubsection{The renormalized 1-loop self-energy in the presence of
$\boldsymbol B$}
%====================================================================

Collecting all terms yields
\begin{equation}
\begin{split}
\Sigma(\pi) &=
\frac{\alpha m}{2\pi}\int_0^\infty\frac{ds}{s}\int_0^1 du\;e^{-isu^2 m^2}
\Bigg\{\frac{e^{-is\Phi}}{\sqrt{\Delta}}\;
\left[ 1+e^{-2i\,Y\sigma^3}+(1-u)e^{-2i\,Y\sigma^3}\;\frac{\pi\!\!\!/}{m}
 \right.\cr
& \hskip 2cm \left.
+(1-u)\left(\frac{1-u}{\Delta}+\frac{u}{\Delta}\;\frac{\sin
Y}{Y}\;e^{-i\, Y\sigma^3}-e^{-2i\, Y\sigma^3}\right)
\frac{\pi\!\!\!/_{\!\perp}}{m}\right]
\cr
& \hskip 4cm\underbrace{
-(1+u)-(\pi\!\!\!/+m)\left[\frac{1-u}{m}-2imu(1-u^2)s\right]
}_{c.t.}
\Bigg\}.
\end{split}
\label{eq:DRstart}
\end{equation}
It is eq.~(3.44) of \cite{DittrichReuter}, which coincides with the
operatorial expression of the
self-energy of an electron in an external $B$ deduced by
Tsai \cite{Tsai1974}

\subsection{Projecting $\boldsymbol{\Sigma(\pi)}$
 on the ``privileged state'':
$\boldsymbol{\delta m}$ for the lowest Landau level}
%sssssssssssssssssssssssssssssssssssssssssssssssssss

The spectrum of a Dirac electron in a pure magnetic field directed along
$z$ is \cite{BerLifPit}
\begin{equation}
\epsilon_n^2 = m^2 + p_z^2 +(2n+1+\sigma_z)\,|e|B,
\end{equation}
in which $\sigma_z=\pm 1$ is $2\; \times$ the spin projection of the electron
on the $z$ axis.
So, at $n=0, \sigma_z=-1, p_z=0$, $\epsilon_n=m$: this so-called
``privileged state'' is nothing more than the lowest Landau level.

We can consider $A_\mu=\left(\begin{array}{c} A_0=0\cr A_x=0\cr A_y=xB\cr
A_z=0\end{array}\right)$ such that  $F_{12}=B$ is the only
non-vanishing component of the classical external $F_{\mu\nu}$.
Then, the wave function of the privileged state of energy $m$ writes
\cite{Luttinger} \cite{KuznetsovMikheev}
\begin{equation}
\psi_{n=0,s=-1,p_y=p_z=0}=\frac{1}{\sqrt{N}}
\left(\frac{|e|B}{\pi}\right)^{1/4}\;e^{-\frac{|e|B}{2}x^2}\;
\left(\begin{array}{c} 0\cr 1 \cr 0 \cr 0\end{array}\right),
\ N\stackrel{\cite{KuznetsovMikheev}}{=} \underbrace{L_y\; L_z}_{dimensions\
along\ y\ and\ z}.
\label{eq:LLL}
\end{equation}

Following (\ref{eq:mdef}), in order to determine  $\delta m$ for the (on
mass-shell) LLL, we shall sandwich the general self-energy operator
(\ref{eq:DRstart}) between two states $|\;\psi>$ defined in (\ref{eq:LLL})
and satisfying $(\pi\!\!\!/ +m)|\;\psi>=0$.

The expression (\ref{eq:DRstart}) involves
$\pi\!\!\!/$ that we shall replace by $-m$, $\Delta$ 
that needs not be transformed, and $\Phi$  which involves
$m^2-\pi\!\!\!/^2$, $\pi_\perp^2$ and $\sigma_{\mu\nu}F^{\mu\nu}$.
The only non-vanishing component of $F^{\mu\nu}$ being $F^{12}=B$,
$\sigma_{\mu\nu}F^{\mu\nu}= \sigma_{12}F^{12}+\sigma_{21}F^{21}=
2\sigma_{12}F^{12}\equiv 2\sigma_3 B$.
Since the electron is an eigenstate of the Dirac equation in the presence
of $B$, $m^2 -\pi\!\!\!/^2$ can be taken to vanish.
$\pi_\perp^2 \equiv \pi_1^2 + \pi_2^2$ is also identical, since the
privileged state has $p_z=0$ and we work in a gauge with $A_z=0$, to $\vec
\pi^2\equiv\pi^2 + \pi_0^2$. One has $\pi\!\!\!/^2 = -\pi^2
+(e/2)\,\sigma_{\mu\nu}F^{\mu\nu}$ such that $\pi_\perp^2 =
-\pi\!\!\!/^2 +\pi_0^2 +\sigma_3\,eB$. Since our gauge for the external $B$
has $A_0=0$, $\pi_0^2 =p_0^2$, which is the energy squared of the electron,
identical to $m^2$ for the privileged state. Therefore, on mass-shell,
$\pi_\perp^2 = \sigma_3\,eB$. 
 When sandwiched between privileged states, \newline
 $<\psi\;|\;\sigma^3\;|\;\psi>=
\left(\begin{array}{cccc}0 & 1 & 0 & 0 \end{array}\right)
 diag(1,-1,1,-1)\left(\begin{array}{c} 0 \cr 1 \cr 0 \cr
0\end{array}\right) = -1$
such that  $\sigma^3$ can be replaced by $(-1)$\newline
 and $\Phi$ shrinks to
$u\,eB\left(1-\frac{\beta}{Y}\right)$.
 $\sigma^3$ can also be replaced by $(-1)$ in the exponentials of
(\ref{eq:DRstart}).

$\Sigma(\pi)$ in (\ref{eq:DRstart}) also involves a term proportional to
 $\pi\!\!\!/_\perp$. Since the privileged state
has $p_z=0$ and we work at $A_z=0$, this is also equal to
$ \vec\gamma.\vec \pi = \gamma^\mu \pi_\mu -\gamma^0 \pi_0
= \pi\!\!\!/ +\gamma^0 p^0$. $<\psi\;|\;\pi\!\!\!/\;|\; \psi>=-m$ such
that\newline
$<\psi\;|\; \pi\!\!\!/_\perp\;|\;\psi> = <\psi\;|\; -m+\gamma^0 p^0\;|\;\psi>$.
Since $\gamma^0=diag(1, 1, -1, -1)$, 
eq.~(\ref{eq:LLL})  yields $<\psi\;|\; \pi\!\!\!/_\perp\;|\;\psi>
=-m+p^0$.
The energy $p^0$ of the privileged state $|\;\psi>$ being equal to $m$, 
this term vanishes.

Gathering all information and simplifications leads finally to
\begin{equation}
\delta m_{LLL}\equiv \Sigma(\pi)_{\pi\!\!\!/+m=0} =
\frac{\alpha m}{2\pi} \int_0^\infty \frac{ds}{s} \int_0^1 du\;
e^{-isu^2 m^2}
\Bigg[\frac{e^{-is\Phi(u,Y)}}{\sqrt{\Delta(u,Y)}}\left(1+ue^{2i Y}\right)
-\underbrace{(1+u)}_{c.t.}\Bigg],
\label{eq:delta-m}
\end{equation}
with $Y = eB su$. $\Delta(u,Y)$ is the same as in (\ref{eq:Delta}),
$\sin\beta$ and $\cos\beta$ the same as in (\ref{eq:beta}).
$\Phi$ in (\ref{eq:Phi}) has shrunk to
\begin{equation}
\Phi(u,Y) =u\,eB\left(1-\frac{\beta(u,Y)}{Y}\right)
= ueB-\frac{\beta(u,Y)}{s}.
\end{equation}
Equivalently
\begin{equation}
\delta m_{LLL} \equiv \Sigma(\pi)_{\pi\!\!\!/+m=0} =
\frac{\alpha m}{2\pi} \int_0^\infty \frac{ds}{s} \int_0^1 du\;
e^{-isu^2m^2}\Bigg[\frac{e^{i[-sueB+\beta(u,Y)]}
+u\,e^{i[sueB+\beta(u,Y)]}}
{\sqrt{\Delta(u,Y)}}-\underbrace{(1+u)}_{c.t.}\Bigg],
\label{eq:deltam}
\end{equation}
which is the expression that we have to evaluate.

\subsection{A few remarks}
%sssssssssssssssssssssssss

*\ At $B\to 0$, $Y\to 0$, $\beta \sim (1-u)Y + {\cal O}(Y^2)$ yields
$\Phi_{B=0}=0$. One also has $\Delta_{B=0}=1$ such that
$\Sigma(\pi)_{B=0}=\displaystyle\frac{\alpha m}{2\pi}\int_0^\infty\frac{ds}{s}\int_0^1
du\;e^{-isu^2m^2}[(1+u)-(1+u)]=0$. This agrees with the renormalization
condition (\ref{eq:renorm}).

*\ $\Delta(u,Y)$, which occurs by its square root, is a seemingly naughty denominator.
Its zeroes $u_\pm$ can be written
$u_+=u_-^\ast = \displaystyle\frac{1-(\sin Y/Y)\;e^{iY}}{\xi(Y)}$, with
$\xi(Y)= 1-2\;\displaystyle\frac{\sin Y \cos Y}{Y}+\left(\displaystyle\frac{\sin Y}{Y}\right)^2$.
The real zeroes $u_+=1=u_-$ are degenerate and are located at $Y=n\pi,
n\not=0$, values at which  $\beta=0$.

*\ The renormalized $\delta m$ given by (\ref{eq:delta-m}) is finite. The
contribution $\propto (1+u)$ from the counterterm is tailored for this.

*\ The (infinite) contribution to $\delta m_{LLL}$ coming from this counterterm
corresponds to its value at $B=0$. It does not  depend  on $B$.
It has been evaluated in pp.~53-56 of \cite{DittrichReuter}:
$\delta m_{B=0} = \lim_{s_0\to 0} \displaystyle\frac{3\alpha m}{4\pi}\left(-\gamma_E +
\ln\displaystyle\frac{1}{im^2 s_0}+\displaystyle\frac56\right)$,
where $s_0$ is the lower limit of
integration for the Schwinger parameter $s_1$ attached to the electron
propagator. It coincides with the result given by
Ritus in \cite{Ritus1975}.

\subsection{Changing variables; the Demeur-Jancovici integral
\cite{Demeur} \cite{Jancovici}}
\label{subsec:changevar}
%ssssssssssssssssssssssssssssssssssssssssssssssssssssssssssssssss

We first perform the change of variables
\begin{equation}
(u,s) \to (u,Y\equiv eBsu) \Rightarrow \frac{du\, ds}{s} = \frac{du\, dY}{Y}.
\label{eq:ch-var}
\end{equation}
In Dittrich-Reuter \cite{DittrichReuter}, $e$ stands for the
(negative) charge of the electron
\footnote{unlike in \cite{Tsai1974} in which, like in
Schwinger, both $q$ and $e$ are introduced. In there, $e$ has the meaning
of the elementary charge $e>0$.}. 
Therefore,  $Y<0$, too, and  $\displaystyle\int_0^\infty \displaystyle\frac{ds}{s} =
\displaystyle\int_0^{-\infty}\displaystyle\frac{dY}{Y}$. Then, $\delta m$ in (\ref{eq:deltam}) becomes
\begin{equation}
\begin{split}
\delta m_{LLL} &= \frac{\alpha m}{2\pi}\int_0^{-\infty} \frac{dY}{Y} \int_0^1
du\;
e^{-iuY\frac{m^2}{eB}}
\Bigg[\frac{e^{i[\beta(u,Y)-Y]}+u\,e^{i[\beta(u,Y)+Y]}}
{\sqrt{\Delta(u,Y)}} -\underbrace{(1+u)}_{from\ c.t.}\Bigg],
\end{split}
\label{eq:delta-m-1}
\end{equation}
which is seen to only depend on $\displaystyle\frac{eB}{m^2}$.
The divergence of $\delta m$ occurs now at $Y\to 0$.
The change (\ref{eq:ch-var}) introduces  a dependence of the
 counterterm on $\displaystyle\frac{|e|B}{m^2}$
\footnote{To summarize in a symbolic (and dirty) way, this change of variables amounts to
rewriting $ \delta m_{LLL}=\left(\infty + \eta(\frac{|e|B}{m^2})\right)
-\infty$ as $\delta m = \left(\infty +\eta(\frac{|e|B}{m^2})
+\zeta(\frac{|e|B}{m^2})\right) - \left(\infty
+\zeta(\frac{|e|B}{m^2})\right)$. $\zeta$ is the dependence on
$\frac{eB}{m^2}$ generated by the change of variables.
We shall then regularize the canceling infinities to get rid of them
and calculate separately $\eta+\zeta$ and $-\zeta$ which give
respectively the $\left(\ln\frac{|e|B}{m^2}\right)^2$ and $\ln
\frac{|e|B}{m^2}$ terms.}.

It is interesting to expand the sole $e^{i\beta}$ into $\cos\beta +
i\sin\beta$, to use the expressions (\ref{eq:delta-m})
 of $\cos\beta$ and $\sin\beta$,  to cast $\delta m$ in the form
\begin{equation}
\delta m_{LLL} = \frac{\alpha m}{2\pi}\int_0^{-\infty} \frac{dY}{Y}
\int_0^1 du\; e^{-iuY\frac{m^2}{eB}}
\left[ (1+u\,e^{2iY})\;\displaystyle\frac{1-u +u\displaystyle\frac{\sin Y}{Y}\,e^{-iY}}{\Delta(u,Y)}-(1+u)
\right]
\end{equation}
and to notice that $\Delta(u,Y) = \Big(1-u +u\displaystyle\frac{\sin
Y}{Y}\,e^{+iY}\Big)\Big(1-u
+u\displaystyle\frac{\sin Y}{Y}\,e^{-iY}\Big)$ to simplify the previous expression into
\begin{equation}
\delta m_{LLL} = \frac{\alpha m}{2\pi}\int_0^{-\infty} \frac{dY}{Y}
\int_0^1 du\; e^{-iuY\frac{m^2}{eB}}
\left[ \displaystyle\frac{1+u\,e^{2iY}}{1-u +u\displaystyle\frac{\sin Y}{Y}\,e^{+iY}}-(1+u)
\right].
\end{equation}
Expressing  $\sin Y$ in the denominator in terms of complex exponentials gives
\begin{equation}
\delta m_{LLL} = \frac{\alpha m}{2\pi}\int_0^{-\infty} dY
\int_0^1 du\; e^{-iuY\frac{m^2}{eB}}
\left[ \frac{2i\left(1+u\,e^{2iY}\right)}{2iY(1-u) +u\left(e^{2iY}-1\right)}
-\frac{1+u}{Y}
\right].
\end{equation}
Going to $t=-iY$ yields
\begin{equation}
\delta m_{LLL} = \frac{\alpha m}{2\pi}\int_0^{+i\infty} dt
\int_0^1 du\; e^{ut\frac{m^2}{eB}}
\left[ \frac{2\left(1+u\,e^{-2t}\right)}{2t(1-u)
+u\left(1-e^{-2t}\right)}
-\frac{1+u}{t}
\right].
\label{eq:ut}
\end{equation}
Last, we change to $z=ut \Rightarrow du\, dt = \frac{du\, dz}{u}$ and get
\begin{equation}
\begin{split}
\delta m_{LLL} &= \frac{\alpha m}{2\pi}\int_0^{+i\infty} dz
\int_0^1 du\; e^{z\frac{m^2}{eB}}
\left[ \frac{2\left(1+u\,e^{-2z/u}\right)}{2z(1-u)
+u^2\left(1-e^{-2z/u}\right)}
-\frac{1+u}{z} \right]\cr
&=\frac{\alpha m}{2\pi}\int_0^{+i\infty} dz
\int_0^1 du\; e^{-z\frac{m^2}{|e|B}}
\left[ \frac{2\left(1+u\,e^{-2z/u}\right)}{2z(1-u)
+u^2\left(1-e^{-2z/u}\right)}
-\frac{1+u}{z} \right].
\end{split}
\label{eq:uz}
\end{equation}
It still differs from eq.\,3 of Jancovici \cite{Jancovici}
 by the 2 following points:\newline
*\  one has $e^{+z\frac{m^2}{eB}}$ instead of
$e^{-z\frac{m^2}{eB}}$ in \cite{Jancovici}; this is due to  $e>0$ in there, while,
here, $e<0$;
\newline
*\ one has $\int_0^{i\infty}dt$ instead of $\int_0^\infty dt$; a Wick
rotation is needed:
$\int_0^{+i\infty} + \int_{1/4\ infinite\ circle} +\int_\infty^0= 2i\pi
\sum residues$. Because of $e^{-z\frac{m^2}{|e|B}}$ the contribution on the
infinite 1/4 circle is vanishing.
That the residue at $z=0$ vanishes is trivial as long as $u$ is not
strictly vanishing. The expansion of the terms between square brackets
in (\ref{eq:uz}) at $z\to 0$ writes indeed
$u-1+ (-\frac53 + \frac{4}{3u} + u) z + \left(-\frac73 - \frac{1}{u^2} +
\frac{7}{3 u} + u\right)z^2+ {\cal O}(z^3)$,
which seemingly displays  poles at $u=0$.
However, without expanding, it also writes, then, $\frac{2}{2z}-\frac{1}{z}=0$,
which shows that the poles at $u=0$ in the expansion at $z\to 0$
are fake and that the residue at $z=0$ always vanishes.
Other poles (we now consider eq.~(\ref{eq:ut})) can only occur when the
denominator of the first term inside brackets vanishes.
That the corresponding $u\stackrel{pole}{=}\frac{2t}{2t+e^{-2t}-1}$ should
be real constrains them  to occur at $t\to in\pi,n\in{\mathbb N}>0$
and $u\to 1$. In general, they satisfy
$2t(1-u)+u(1-e^{-2t})=0$ which, setting $t=t_1+it_2, t_1,t_2 \in{\mathbb
R}$,
yields the  2 equations $e^{-2t_1}\;\cos 2t_2 = 1+2\eta t_1,\ 
e^{-2t_1}\;\sin 2t_2 = -2\eta t_2,\ \eta=\frac{1-u}{u}\geq0$.
Since $t_1\to 0$, one may expand the first relation at this limit, which
yields $\cos 2t_2 -1 = 2t_1(\eta+\cos 2t_2)$. As $t_2 \to n\pi$, $\cos 2t_2
>0$ and $\cos 2t_2 -1<0$, which, since $\eta>0$, constrains $t_1$ to stay negative
\footnote{The 2nd relation then tells us that $\sin 2t_2 <0$, which means
that the poles correspond to $t_2=n\pi-\epsilon,\epsilon >0$.}
. Therefore, the potentially troublesome poles lie in reality
on the left of the imaginary $t$ axis along which the integration is done
and should not be accounted for when doing a Wick rotation. It gives
\begin{equation}
\begin{split}
\delta m_{LLL} &=
\frac{\alpha m}{4\pi}\; 2\int_0^{\infty} dz
\int_0^1 du\; e^{-z\frac{m^2}{|e|B}}
\Bigg[ \frac{2\left(1+u\,e^{-2z/u}\right)}{2z(1-u)
+u^2\left(1-e^{-2z/u}\right)}
-\underbrace{\frac{1+u}{z}}_{from\ c.t.} \Bigg].%\cr
%&= \frac{\alpha m}{4\pi}\;I(L),\quad L=\frac{(\hbar)|e|B}{(c^3)m^2}
\end{split}
\label{eq:SchJanc}
\end{equation}
(\ref{eq:SchJanc}) is now the same as Jancovici's eq.\,3 \cite{Jancovici}
(see eqs.~(\ref{eq:Iex},\ref{eq:deltamJ}) below).
This proves in particular that the latter (and therefore Demeur's
calculation \cite{Demeur}) satisfy the same ``on mass shell'' renormalization
conditions (\ref{eq:renorm}), which was not clear in \cite{Demeur}.

%SSSSSSSSSSSSSSSSSSSSSSSSSSSSSSSSSSSSSSSSSSSSSSSSSSSSSSSSSSSSSSSSSSSSSSSSS
\section{Calculating  Jancovici's integral \cite{Jancovici}} % appro2.nb
%SSSSSSSSSSSSSSSSSSSSSSSSSSSSSSSSSSSSSSSSSSSSSSSSSSSSSSSSSSSSSSSSSSSSSSSSS

\subsection{Generalities and definition}
%sssssssssssssssssssssssssssssssssssssss

Along with Jancovici \cite{Jancovici}, let us write the rest energy of the electron 
\begin{equation}
E_0 = m(c^2)\left(1+\frac{\alpha}{4\pi}\;I(L)\right),
\quad L=\frac{(\hbar)|e|B}{(c^3)m^2}
\end{equation}
in which, at all orders in $B$
\begin{equation}
\begin{split}
& I(L) =2\int_0^\infty dz\;e^{-z/L}\int_0^1 dv\left(
\frac{2\left(1+v\,e^{-2z/v}\right)}{2z(1-v)+v^2\left(1-e^{-2z/v}\right)}
-\frac{1+v}{z}
\right) = 2\int_0^\infty dz\;e^{-z/L}\int_0^1 dv\;f(v,z),\cr
& \hskip 3cm f(v,z)=
\frac{2\left(1+v\,e^{-2z/v}\right)}{2z(1-v)+v^2\left(1-e^{-2z/v}\right)}
-\frac{1+v}{z}.
\end{split}
\label{eq:Iex}
\end{equation}
Jancovici \cite{Jancovici} defines accordingly (we set hereafter $\hbar=1=c$)
\begin{equation}
\delta m = \frac{\alpha m}{4\pi}\; I(L)
\label{eq:deltamJ}
\end{equation}
such that the $I(L)$ in (\ref{eq:Iex}) coincides  with the one in
(\ref{eq:SchJanc}). Since the same external electron states are concerned
in the 2 calculations, we have proved that they are equivalent.

$I(L)$ has been obtained from Demeur's original integral \cite{Demeur}
\footnote{It is eq.~(21) of \textsection\,8: ``La self-\'energie de
l'\'electron'', p.~78 of \cite{Demeur}.}
\footnote{It has  been manifestly obtained with an internal photon
in the Feynman gauge (see eq.~(1) p.~56 of \cite{Demeur}).}
\begin{equation}
D(L)=\int_0^1 dv\;(1+v)\int\frac{dw}{w}\;\frac{w}{|w|}\;e^{ivw}\;
\frac{2iLw(v\,e^{2iLw}+1)}{(1+v)[v\,e^{2iLw}+2iLw(1-v)-v]}
\label{eq:Demeur}
\end{equation}
by subtracting its value  at $B=0 \Leftrightarrow L=0$
and after the change of variables $z=-iLvw$.
Therefore, (\ref{eq:deltamJ}) corresponds to the magnetic radiative corrections
to the electron mass, after subtracting the self-energy of the ``free''
(i.e. at $B=0$) electron
\footnote{See Demeur \cite{Demeur} chapitre III ``Les corrections
radiatives magn\'etiques'', \textsection\,1 ``La self-\'energie'', p.55}. 
The latter corresponds to the term $\propto \frac{1+v}{z}$ in the
integrand of (\ref{eq:Iex}). Accordingly, (\ref{eq:deltamJ}) satisfies
$\delta m \stackrel{B\to 0}{\to}0$.

\subsection{First steps: a simple convergent approximation for
$\boldsymbol{L\equiv eB/m^2>75}$}
\label{subsec:first-steps}
%ssssssssssssssssssssssssssssssssssssssssssssssssssssssssssssss

We want an analytical expression for $I(L)$ valid for large values
of the magnetic field, say $\displaystyle\frac{|e|B}{m^2}>75$.
That $I(L)$ can easily be integrated numerically makes checks easy.

The two integrals making up (\ref{eq:Iex}) both diverge at $z\to 0$. 
The cancellation of the  divergences is ensured by the first renormalization
condition (\ref{eq:renorm}), but its practical implementation needs  a
regularization.

Following Jancovici \cite{Jancovici}, one splits $I(L)$ into
  $\int _0^\infty dz = \int_0^a dz + \int _a^\infty dz$,
with $a$ large enough such that $e^{-2z/v}\ll 1$ can be neglected
inside $f(v,z)$.  Since $v
\in [0,1]$, this requires at least $a\geq 1$, that we check numerically.
$I(L)$ can then be approximated by
\begin{equation}
I(L)\approx 2\int_0^a dz\;e^{-z/L}\int_0^1 dv\;
f(v,z)
+2 \int_a^\infty dz\;e^{-z/L}\int_0^1
dv\;\left(\frac{2}{v^2+2z(1-v)}-\frac{1+v}{z}\right),
\label{eq:Iap}
\end{equation}
in which the second integral is manifestly convergent. We focus on the
first one, which includes the two canceling divergences.
It turns out, as in \cite{Jancovici}, that, for $L$ large enough, for example
$L>75$, its numerical value decreases with $a$ and that one can go very safely down to
$a=1$ at which  it is totally negligible with respect to the value of the
full $I$
\footnote{We proceed as follows. Though $f(0,z)=0$,
$f(v,z)$ cannot be integrated $\int_0^1 dv$ at small $z$ because, as
already mentioned in subsection \ref{subsec:changevar}, its expansion
has (fake) poles at $v=0$ and numerical integration becomes itself hazardous.
To achieve it safely, 
we regularize the first integral in (\ref{eq:Iap}) by introducing a small parameter
$\epsilon$,  replace $\int_0^1 dv\; f(v,z)$ with $\int_\epsilon^1
dv\;f(v,z)$, then decrease $\epsilon = 10^{-3}, 10^{-6}, 10^{-9} \ldots$
while checking stability.}.
We thus approximate, for $L\geq 75$
\begin{equation}
I(L)\stackrel{L\geq 75}{\approx}
2 \int_{a=1}^\infty dz\;e^{-z/L}\int_0^1
dv\;\left(\frac{2}{v^2+2z(1-v)}-\frac{1+v}{z}\right).
\label{eq:Iap2}
\end{equation}
The second contribution to (\ref{eq:Iap2}), which comes from the
counterterm
\footnote{This term  was neglected in
eq.~(4) of \cite{Jancovici}, where only  $\ln^2$  are focused on.}
, is easily integrated, and one gets
\begin{equation}
I(L)\stackrel{L\geq 75}{\approx}2
\underbrace{ \int_1^\infty
dz\;e^{-z/L}\int_0^1
dv\;\frac{2}{v^2+2z(1-v)}}_{J(L)}
\quad -3\; \Gamma(0,1/L)
=2\,J(L) -3\; \Gamma(0,1/L)
\label{eq:Iap3}
\end{equation}
in which $\Gamma(0,z)$ is the incomplete Gamma function $\Gamma(0,z) =
\displaystyle\int_z^\infty \displaystyle\frac{e^{-t}}{t}\;dt$. The integral
$\displaystyle\int_0^1
dv\;\displaystyle\frac{2}{v^2+2z(1-v)}$ can be easily performed analytically, too, leading to
\begin{equation}
I(L)\stackrel{L\geq 75}{\approx}
2\underbrace{\int_1^\infty dz\;e^{-z/L}\;
\frac{\ln\left(z-1+\sqrt{z(z-2)}\right)}{\sqrt{z(z-2)}}}_{J(L)}
\quad-\underbrace{3\; \Gamma(0,1/L)}_{from\ c.t.}.
\label{eq:Iapp2}
\end{equation}
The result of  the change of variables done in
subsection \ref{subsec:changevar} associated with the regularization-approximation
just performed is a sum of two finite integrals. The most
peculiar and also the most important for our purposes is the second one
which originates from the counter-term and includes the large
$\ln\displaystyle\frac{|e|B}{m^2}$ generally ignored. Its occurrence 
is non-trivial and only appears through the change of variables
(\ref{eq:ch-var}).

\subsection{Further evaluation}
\label{subsec:Japprox}
%ssssssssssssssssssssssssssssss

$J(L)\equiv \displaystyle\int_1^\infty dz\;e^{-z/L}\;g(z),\ g(z) =
\displaystyle\frac{\ln\left(z-1+\sqrt{z(z-2)}\right)}{\sqrt{z(z-2)}}$
cannot be integrated exactly but, following \cite{Vysotsky2010},
 one can find an accurate approximation for the integrand
\begin{equation}
g_{app}(z) \approx \frac{\ln z}{z} + \frac{\pi}{2}\;\frac{1}{z^\beta},\quad
\beta=1.175
\end{equation}
as shown on Fig.~2 below where the 2 curves for the exact $g$ (blue)
and the approximate $g_{app}$ (yellow) are practically indistinguishable.
%
%ffffffffffffffffffffffffff FIGURE 2 g.pdf fffffffffffffffffffffffffffffff
%\vbox{
\begin{center}
\includegraphics[width=6cm, height=4cm]{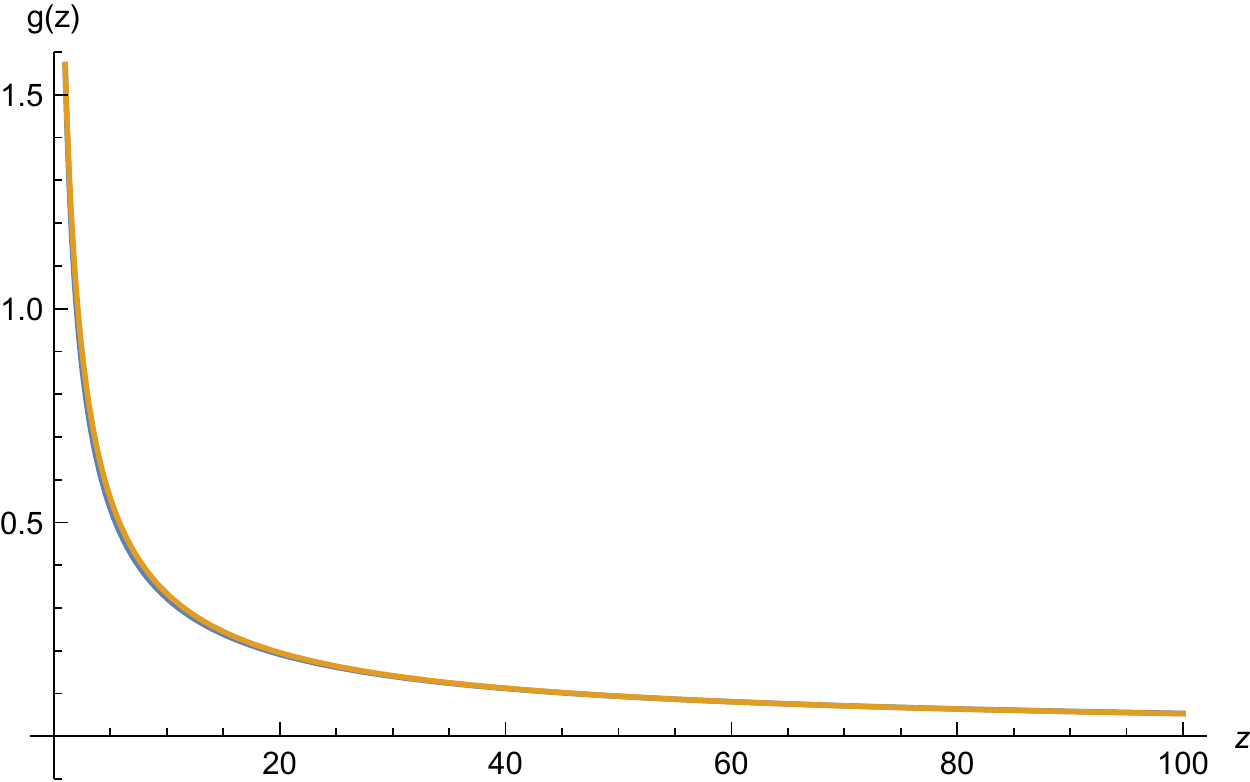}

{\em Fig.~2: exact (blue) and approximate (yellow) values for the integrand
$g(z)$ of  $J(L)$.}
\end{center}
%}
%ffffffffffffffffffffffffff FIGURE 2 g.pdf fffffffffffffffffffffffffffffff
%
Without the term $\displaystyle\frac{\pi}{2}\;\displaystyle\frac{1}{z^\beta}$,
$g$ would go to $0$ instead
of $\displaystyle\frac{\pi}{2}$ at $z=1$. This term yields in particular the term
$\propto \displaystyle\frac{1}{L^{\beta -1}}$ in the expansion of $J_{app}$ at $L \to
\infty$.
The integration can now be done analytically, leading to
\begin{equation}
J_{app}(L)=\int_1^\infty dz\;e^{-z/L}\;\left(
\frac{\ln z}{z} + \frac{\pi}{2}\;\frac{1}{z^\beta}\right)=
\underbrace{\frac{\pi}{2}\, \text{ExpIntegralE}\left[\beta,
\frac{1}{L}\right]}_{from\
\frac{\pi}{2}\frac{1}{z^\beta}} +
 \underbrace{\text{MeijerG}\left[\{(\ ), (1, 1)\}, \{(0, 0, 0), (\ )\},
\frac{1}{L}\right]}_{from\ \frac{\ln z}{z}}.
\end{equation}
We compare in Fig.~3 the integrals $J(L)$ (blue) and $J_{app}(L)$ (yellow),
which prove extremely close to each other.
%
%ffffffffffffffffffffffffff FIGURE 3 J.pdf fffffffffffffffffffffffffffffff
\begin{center}
\includegraphics[width=6cm, height=4cm]{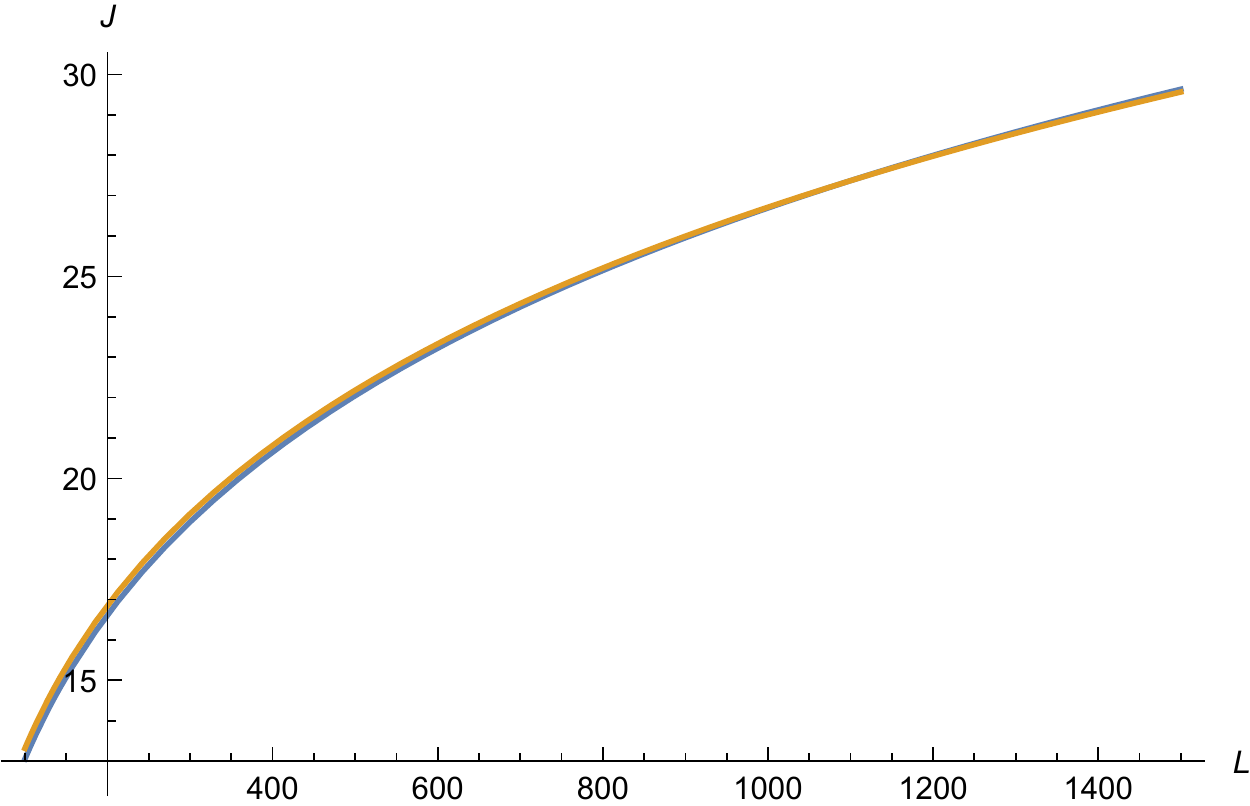}

{\em Fig.~3: exact (blue) and approximate (yellow) values for   $J(L)$.}
\end{center}
%ffffffffffffffffffffffffff FIGURE 3 J.pdf fffffffffffffffffffffffffffffff
%

\subsection{Final result}
%ssssssssssssssssssssssss

The final result is obtained by expanding $J_{app}(L)$ and $\Gamma(0,1/L)$
at large $L$
\begin{equation}
\begin{split}
& \hskip -1cm J_{app} \stackrel{L\to\infty}{\simeq}
\frac{1}{L^\beta}\left(
\frac{\pi}{2} L\,\Gamma[1 - \beta]+{\cal O}(\frac{1}{L^2})
\right)
+\frac{\gamma_E^2}{2}
+\frac{\pi}{12}\left(\frac{6}{\beta-1}+\pi \right)
-\frac12\,\ln L\left(2\gamma_E -\ln L\right)
+\frac{-1 + \frac{\pi}{4 - 2 \beta}}{L}
+{\cal O}(\frac{1}{L^2}),\cr
& \hskip 1cm \Gamma(0,1/L) \stackrel{L\to\infty}{\simeq} -\gamma_E +\ln L +\frac{1}{L}
+{\cal O}(\frac{1}{L^2})\quad \text{(comes from the counterterm)}
\end{split}
\label{eq:Iapp3}
\end{equation}
which yields for $I(L)$ written in (\ref{eq:Iapp2})
\begin{equation}
\begin{split}
I_{app}(L,\beta) & \stackrel{L\geq 75}{\approx}
 \gamma_E^2 \underbrace{+3\gamma_E}_{from\ c.t.} +\frac{\pi}{\beta-1}
+\frac{\pi^2}{6}
+\frac{\pi\;\Gamma[1-\beta]}{L^{\beta-1}}
-\ln L\Big(2\gamma_E+\underbrace{3}_{from\ c.t.}\Big)
+ \boldsymbol{(\ln L)^2}\cr
& \hskip 5cm
+\frac{1}{L}\Big(\frac{\pi}{2-\beta}-2\underbrace{-3}_{from\ c.t.}\Big)
+{\cal O}(\frac{1}{L^{\geq 2}})\cr
&= \left(\ln L-\gamma_E-\frac32\right)^2 -\frac94  +\frac{\pi}{\beta-1}
+\frac{\pi^2}{6} +\frac{\pi\;\Gamma[1-\beta]}{L^{\beta-1}}
+\frac{1}{L}\left(\frac{\pi}{2-\beta}-5\right)
+{\cal O}(\frac{1}{L^{\geq 2}}).
\end{split}
\label{eq:Iapp4}
\end{equation}
The terms under-braced ``from\ c.t.'' result from the subtraction of the
electron self-energy at $B=0$;
they include a large  $-3(\ln L-\gamma_E)$, which therefore originates from the
counterterm (together with part of the constant term in $\delta m$).

At $L\geq 75$ the term $\propto 1/L$ can be very safely neglected
and one can approximate
\begin{equation}
I_{app}(L,\beta)\stackrel{L\geq 75}{\approx}
\left(\ln L-\gamma_E-\frac32\right)^2 -\frac94
 +\frac{\pi}{\beta-1} +\frac{\pi^2}{6}
+\frac{\pi\;\Gamma[1-\beta]}{L^{\beta-1}}
+{\cal O}(\frac{1}{L}),\quad \beta \approx 1.175
\label{eq:Iapp5}
\end{equation}
which is very different, as we shall see,
 from the brutal approximation $I_{app}\approx (\ln L)^2$ that has been
systematically used in the following years. At $\beta = 1.175$, one gets
explicitly
\begin{equation}
I_{app}(L,\beta =1.175) \stackrel{L\geq 75}{\approx}
\ln L (\ln L -4.15443)
- \frac{20.4164}{L^{0.175}} +
21.6617 %- \frac{1.19201}{L}
+{\cal O}(\frac{1}{L^{\geq 1}}).
\label{eq:Iapp6}
\end{equation}
We plot in Fig.~4 the different contributions to the Demeur-Jancovici
integral: the green curve is the
constant term, the yellow one is the inverse power, the brown one the $\ln$
contribution, the red one the $(\ln)^2$, and the blue curve is the global
result.
The comparison between the red and blue curve is that between the
systematically used  $(\ln)^2$ approximation and our accurate evaluation
(\ref{eq:Iapp5}).
A large cancellation between $(\ln)^2$ and $\ln$ terms
\footnote{They  exactly cancel at $\ln L \approx 4.15443
\Leftrightarrow B \approx 63\,B_0$, where $B_0 \equiv \frac{m^2}{|e|}$
 is the ``Schwinger critical field''.}
 makes in particular the role of the large constant important.
%
%ffffffffffffffffffffffffff FIGURE 4 contribs1.pdf contribs.pdf ffffffff
\begin{center}
\includegraphics[width=6cm, height=4cm]{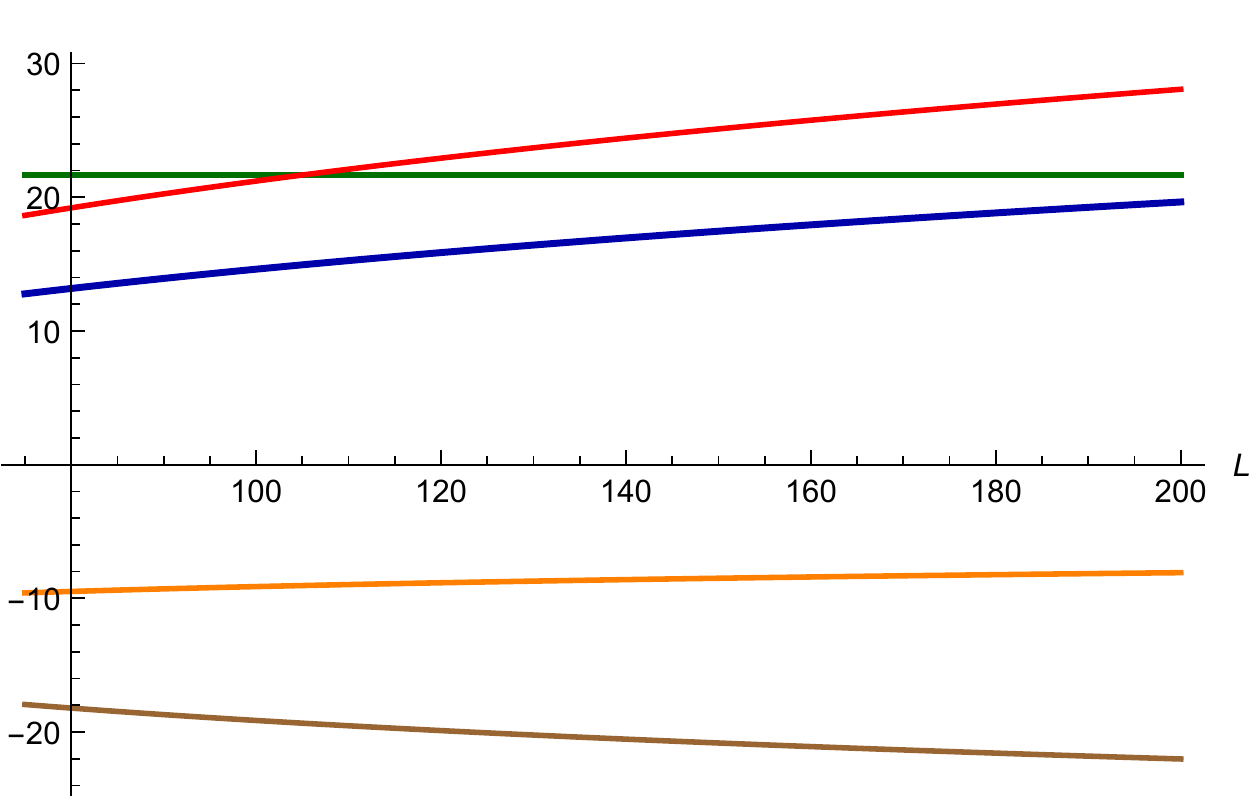}
\hskip 2cm
\includegraphics[width=6cm, height=4cm]{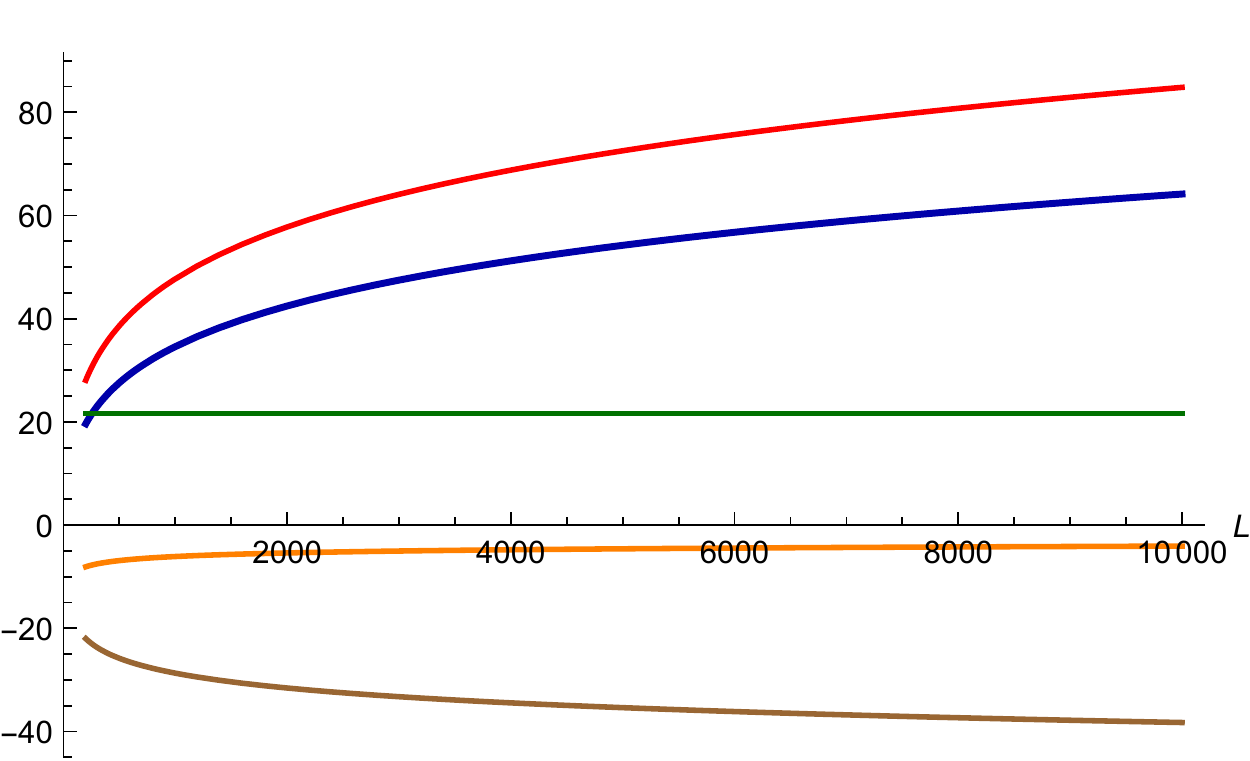}

{\em Fig.~4:  contributions to the Demeur-Jancovici integral;
constant term (green), inverse power (yellow), $\ln$ (brown), $\ln^2$
(red), sum of all (blue). }
\end{center}
%ffffffffffffffffffffffffff FIGURE 4 contribs1.pdf contribs.pdf ffffffff
%
The $(\ln L)^2$ exceeds by $45\%$ the precise estimate at $L=100$ and still by
$32\%$ at $L=10000$.
These values of $L$ correspond to already gigantic
magnetic fields that cannot be produced on earth (hundred times the
Schwinger ``critical'' $B_c$).
The absolute difference increases with $L$ while the
relative difference decreases very slowly.
One needs $L> 2\times 10^{17}$ for the relative error to be 
smaller than $1/10$, which is  a totally unrealistic
value of $B$.

Jancovici mentioned at the end of his work \cite{Jancovici} a
refined estimate
$I(L) \simeq \Big(\ln 2L-\gamma_E-\displaystyle\frac32\Big)^2 + A$ with  $-6 \leq A \leq +7$.
Actually, the value $A=3.5$ yields a good agreement  with our calculation in
the range $75 \leq L \leq 100\,000$, as shown in Figs.~5. It corresponds to
$
I(L)_{Jancovici} \approx (\ln L)^2 -1.768\,\ln L + 5.416.
$
%
%ffffffffffffffffffffffffff FIGURE 5 compar-1.pdf compar-2.pdf fffffffff
\begin{center}
\includegraphics[width=6cm, height=4cm]{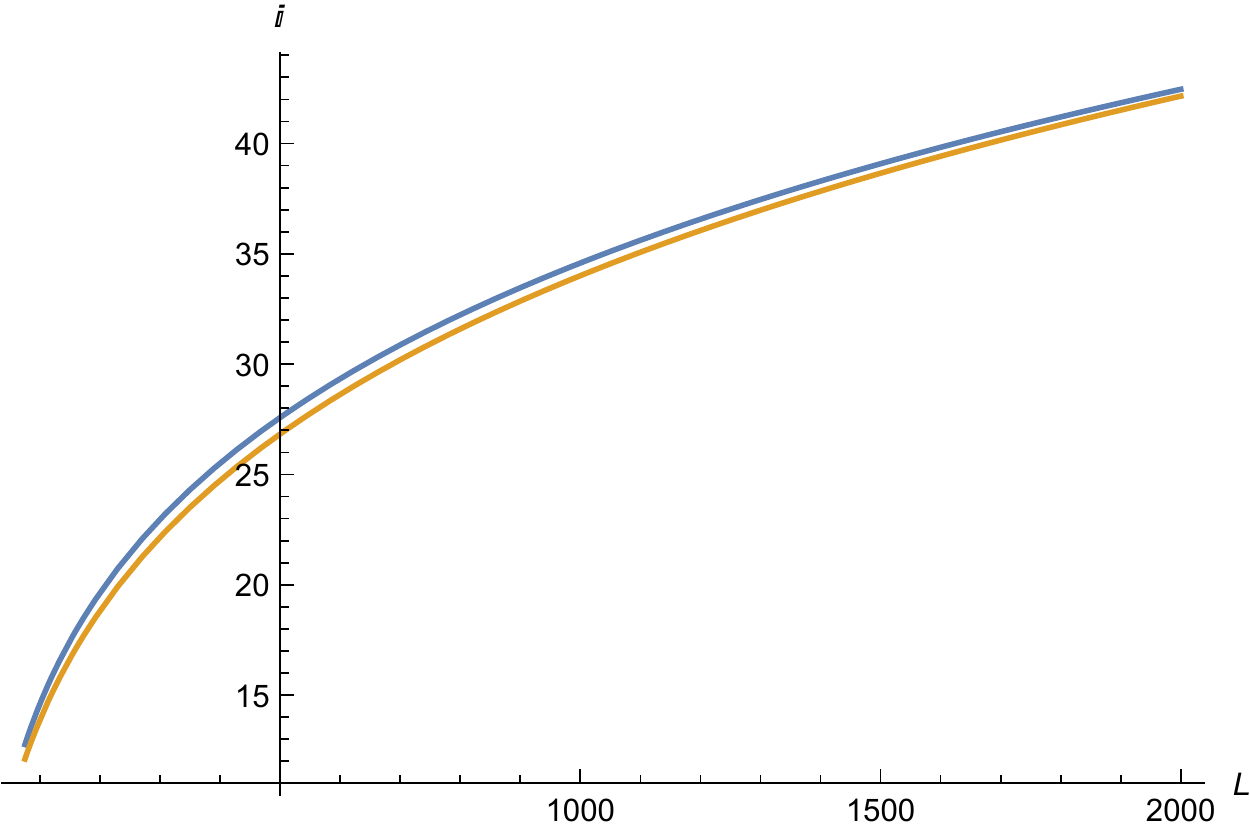}
\hskip 2cm
\includegraphics[width=6cm, height=4cm]{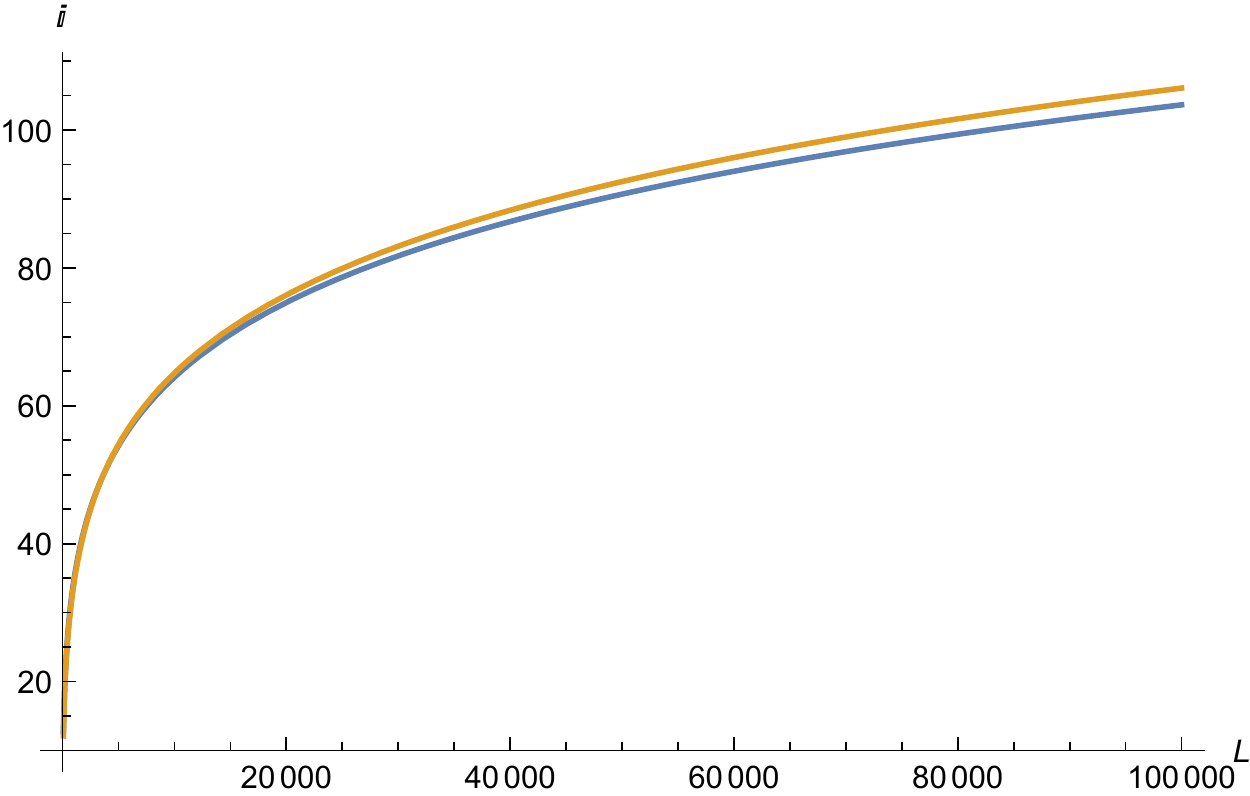}

{\em Fig.~5:  comparison between the present calculation
 of $I(L)$ (blue) and
Jancovici's final refined estimate with $A=3.5$, $I(L)\simeq(\ln
2L-\gamma_E-\frac32)^2 + 3.5$  (yellow).}
\end{center}
%ffffffffffffffffffffffffff FIGURE 5 compar-1.pdf compar-2.pdf fffffffff
%
A comparison is due between the present calculation (\ref{eq:Iapp6}) and
Jancovici's, in particular because the former involves $(\ln L + \ldots)^2$ as
seen in (\ref{eq:Iapp4}) while the latter involves $(\ln 2L+\ldots)^2$.
The result is that, though being very close numerically, the former
includes, in addition to the $\ln^2$, large canceling ($\ln$, constant and inverse power) 
contributions,   while the latter  includes  smaller log,
 constant and no inverse power. This could raise questions about which
evaluation is closer to the exact result. However, the accuracy of
the ``analytical approximation''  to $J(L)$ that we performed in subsection
\ref{subsec:Japprox} and the fact  that it is hard to know how Jancovici
got his ``tedious but straightforward'' \cite{Jancovici}
 estimate tend to support our calculation and the presence, in particular,
 of a large single logarithm.

%SSSSSSSSSSSSSSSSSSSSSSSSSSSSSSSSSSSSSSSSSSS
\section{Concluding remarks and  challenges}
%SSSSSSSSSSSSSSSSSSSSSSSSSSSSSSSSSSSSSSSSSSS

In view of these results, it appears illegitimate to
approximate the integral of Demeur-Jancovici (and the corresponding $\delta
m$  of the electron at 1-loop) by the sole term proportional to
 $\Big(\ln\displaystyle\frac{|e|B}{m^2}\Big)^2$.
Still, all formal
manipulations that have been made until recently, like resummations at a
higher number of loops of a certain class of diagrams, have only
concerned  the double log contribution and its eventual later shrinking
 to a single log by the modification of the photonic vacuum polarization
\footnote{The screening of the Coulomb potential and effective mass for the
photon are obtained by resumming the geometric series of vacuum
polarizations at 1-loop (see for example \cite{Vysotsky2010}).
 Consistency forbids therefore that a massive
photon be inserted into the 1-loop self-energy of the electron.}.

The stakes for improvements are twofold:
include large corrections and fulfill suitable renormalization conditions.
They  are obviously technically challenging, since
the manipulations mentioned above should  be generalized
beyond the ``leading (double-)log approximation''. 
Achieving a reliable resummation at a large number of loops
is  all the more non-trivial as it furthermore needs to satisfy at
each order the appropriate renormalization conditions, that, as we have
seen, control in particular the large single logarithm.
This however  becomes necessary when going to
very large values of $\ln\displaystyle\frac{|e|B}{m^2}$
or when considering theories more strongly coupled than standard QED.

A second obvious challenge is to extend the present calculation to
higher Landau levels of the external electrons.

Though it is premature to make any prospect, the sharp damping of
 $\delta m$  with respect to previous approximations that we have found at
1-loop  cannot but suggest that physical consequences could go along the
same way.
This is left for later investigations.

%****************************************************************

\vskip 5mm

\underline{\em Acknowledgments:} 
it is a pleasure to thank M.~Bellon and J.B.~Zuber for helping me to improve
this work.

%****************************************************************
\newpage
%\vskip 10mm
%****************************************************************

%****************************************************************

\end{document}